\documentclass{iopart}
\usepackage{iopams}
\usepackage{amssymb,amstext,amsfonts}
\usepackage[final]{graphicx}
\usepackage[sort&compress,comma,square,numbers]{natbib}

\newlength\figurewidth
\AtBeginDocument{\setlength\figurewidth{.65\linewidth}}
\let\bs\boldsymbol

\usepackage{marginnote}
\usepackage{color}

\let\bs\boldsymbol
\def\kT{\ensuremath{k_\text{B}T}}

\newenvironment{subequations}{}{}

\def\eqref#1{(\ref{#1})}

\def\uer{
  Institut f\"ur Theoretische Physik, Universit\"at Erlangen-N\"urnberg,
  Staudtstra\ss{}e~7, 91058 Erlangen, Germany}
\def\dlr{
  Institut f\"ur Materialphysik im Weltraum,
  Deutsches Zentrum f\"ur Luft- und Raumfahrt (DLR),
  51170 K\"oln, Germany}
\def\ukn{
  Fachbereich Physik, Universit\"at Konstanz, 78457 Konstanz, Germany}
\def\zkk{
  Zukunftskolleg, Universit\"at Konstanz, 78457 Konstanz, Germany}
\def\mpi{
  Max-Planck-Institut f\"ur Metallforschung, Heisenbergstra\ss{}e~3,
  70569 Stuttgart and Institut f\"ur Theoretische und Angewandte Physik,
  Universit\"at Stuttgart, Pfaffenwaldring~57, 70569 Stuttgart, Germany}

\begin{document}
\title[Long-Wavelength Anomalies of MCT]{Long-Wavelength Anomalies in the Asymptotic Behavior of Mode-Coupling
  Theory}
\author{S~K~Schnyder,$^1$ F~H\"of{}ling,$^2$ T~Franosch,$^3$ and Th~Voigtmann$^{1,4}$}
\address{%
  $^1$\dlr\ and \ukn\\
  $^2$\mpi\\
  $^3$\uer\\
  $^4$\zkk}
\date{\today}

\begin{abstract}
We discuss the dynamic behavior of a tagged particle close to a
classical localization transition in the framework of the mode-coupling
theory of the glass transition.
Asymptotic results are derived for the order parameter as well as the
dynamic correlation functions and the mean-squared displacement close
to the transition. The influence of an infrared cutoff is discussed.
\end{abstract}


\section{Introduction}

Recently, there has been renewed interest in the slow transport of
particles in heterogeneous disordered environments, for example, the diffusion
in porous media
\cite{Kurzidim09,Kim09,Hoefling06,Hoefling07,Hoefling08},
or anomalous-diffusion phenomena
in binary glass forming mixtures \cite{Voigtmann09,Moreno06,Moreno06b}.
Applications arise in many diverse subjects of science and
engineering, notably the physics of
transport in crowded biological systems \cite{Hoefling10pre},
ion transport in glass formers \cite{Bunde98,Voigtmann06},
or the geophysics of volcano eruptions \cite{Dingwell96}.

One way of modeling such systems is by binary systems composed of a
slowly relaxing ``matrix'' component and a ``fast'' species responsible
for the transport. This naturally leads to binary mixtures of
particles with disparate sizes, be they soft \cite{Voigtmann09} or hard spheres
\cite{Voigtmann10pre,Hajnal09}.
If the slow matrix is driven to glassy arrest, it
forms a random heterogeneous background that can be considered as
frozen on the time scale of fast-particle transport. One can also
freeze the matrix from the outset, and is led to
quenched-annealed mixtures
\cite{Krakoviack05,Krakoviack07,Krakoviack09} or, abstracting
even more, the Lorentz gas (LG) model \cite{Hoefling06,Hoefling07,Hoefling08},
a single mobile particle in a random disordered matrix.

All these model systems show intriguing transport phenomena when the
matrix is dense enough, although the precise connection for example between
the LG model and glass-forming binary mixtures remains to be established
\cite{confit}. Most notably, the mobile species can undergo a
localization transition, where its long-range transport ceases because the
embedding matrix is too dense. For the LG, it is understood that this
transition is a dynamic critical phenomenon \cite{Havlin2002} associated with
the percolation of void space between the frozen obstacles. This leads to
distinct scaling predictions close to the transition, some of which have
only been worked out recently \cite{Kammerer08}.

One of these predictions is anomalous power-law diffusion: the mean-squared
displacement (MSD) of the mobile particle, $\delta r^2(t)$, for $t\to\infty$
does not grow linearly in time as expected for ordinary diffusion.
Instead, $\delta r^2(t\to\infty)\sim t^{2/z}$ with a dynamic critical
exponent $z>2$ is predicted and verified in extensive computer
simulations for the LG \cite{Hoefling06}. In three dimensions, $z=6.25$
is established, so that $\delta r^2(t\to\infty)\sim t^{0.32}$.
It is not yet clear, whether this asymptotic prediction also holds for
a binary mixture, where correlations and residual thermal motion of the
background matrix, and the excluded-volume interaction of many tracers
come into play \cite{Voigtmann09}.

For dense, glass-forming binary mixtures on the other hand, the
mode-coupling theory of the glass transition (MCT) \cite{Goetze.2009}
provides an excellent theoretical framework that allows many detailed
predictions \cite{Goetze03,Hajnal09,Voigtmann10pre}.
The glass transition in this framework comes about because of collective local
nearest-neighbor caging; quite unlike the single-particle dynamic
critical phenomenon that drives the localization transition in the LG model
involving a divergent length scale.
Connected to this, the glass transition is a discontinuous dynamic
transition -- the long-time limit (called nonergodicity factor)
of the collective dynamic density
correlation functions jumps from zero to a nonzero value at the glass
transition. In contrast, the localization transition is continuous in this
sense and it belongs to a universality class of its own \cite{Franosch94}.
Theories along the lines of the current MCT have been applied to the
LG previously \cite{Goetze81b,Goetze81c,Goetze82},
and the appearance of continuous transitions
in MCT applied to both quenched-annealed
\cite{Krakoviack05,Krakoviack07,Krakoviack09} and
size-disparate mixtures \cite{Voigtmann10pre,Hajnal09} cast new attention to
the description of such divergent-length localization transitions within
that theory.

Most previous studies of localization transitions within MCT
have put the influence of a divergent length scale on the theory's vertex
out of focus, explicitly dropping dependencies on wave number $q$
-- in the memory kernel only \cite{Goetze81b}, or fully in the
form of a schematic model \cite{Sjoegren86} --
or by silently introducing small-$q$
cutoffs through numerical discretization.
In fact, such a cutoff can be argued for since the small-$q$ structure
of MCT's vertex is incorrect at least in the LG \cite{Leutheusser1983}.
However, it seems to be largely undocumented so far how sensitive the
expected results are to such a cutoff.

Here we document the dependence of asymptotic results of MCT close to
localization transitions on the small-wave-number regime. The analytic
results are corroborated by careful numerical evaluation of the MCT
integrals for the tagged-particle correlators. We identify the regime
where the solutions are dominated by the small-$q$ features by studying
the dependence of the solutions on an ad-hoc infrared cutoff.
For simplicity, we obtain numerical
solutions for the LG model, but the generic picture is the same
in the MCT for mixtures or quenched-annealed systems.

\section{Methods}\label{methods}

Consider a mixture of classical hard-sphere particles in $d$ dimensions,
with diameters $d_\alpha$, where Greek indices label the different
species. $N=\sum_\alpha N_\alpha$ is the total number of particles in
a box of volume $V$, by $x_\alpha=N_\alpha/V$ we denote the respective number
concentrations. Since we are interested in single-particle transport
quantities, we consider the
dynamic tagged-particle density correlation function,
\begin{equation}
  \phi_\alpha(q,t)=\left\langle\varrho_\alpha^{s*}(\vec q,t)
  \varrho_\alpha^s(\vec q)\right\rangle\,,
\end{equation}
where $\varrho_\alpha^s(\vec q,t)=\exp[i\vec q\cdot\vec r_{\alpha,s}(t)]$
is the corresponding tagged-particle density fluctuation to wave vector
$\vec q$, and $\vec r_{\alpha,s}(t)$ the trajectory of the singled-out
tracer particle (taken to be of species $\alpha$). $\phi_\alpha(q,t)$
is real-valued and does not depend on the direction of the wave vector $\vec q$
for a translational-invariant and isotropic system; we also omit the $t=0$
argument since we assume all dynamics to be time-translational invariant.
For convenience, it serves to introduce the Laplace transformed
$
  \phi_\alpha(q,z)=i\int_0^\infty e^{izt}\phi_\alpha(q,t)\,dt
$
for complex frequency $z$ such that the integral converges.

Starting from Newton's equations of motion,
the time evolution of the tagged-particle correlation function can be
represented by a Mori-Zwanzig equation
\cite{Goetze.2009},
\begin{subequations}
\begin{equation}\label{phisz}
  \phi_\alpha(q,z)=-\left[z
  -\frac{q^2}{z/v_{\text{th},\alpha}^2+i\nu_{\alpha}q^2+m_\alpha(q,z)}
  \right]^{-1}\,,
\end{equation}
where $v_{\text{th},\alpha}=\sqrt{\kT/m_\alpha}$ is the thermal velocity
of the particles with mass $m_\alpha$, characterizing the ballistic
short-time motion.
The term $\nu_\alpha=\sqrt\pi\varrho v_{\text{th},\alpha}\sum_\beta
x_\beta g_{\alpha\beta}R_{\alpha\beta}^2\sqrt{2\mu_{\alpha\beta}}$
arises because for hard spheres, the
dynamics is described by a non-self-adjoint pseudo-Liouville operator
\cite{Resibois.1977}, and is a central quantity calculated in
Enskog theory \cite{BoonYip,Foffi.2003}. For brevity, we have set
$R_{\alpha\beta}=d_\alpha+d_\beta$ and
$\mu_{\alpha\beta}=m_\beta/(m_\alpha+m_\beta)$, and denoted by
$g_{\alpha\beta}$ the value of the radial distribution function
for two particles of species $\alpha$ and $\beta$ at contact.
In the time domain, Eq.~\eqref{phisz} reads
\begin{eqnarray}\label{phist}
  \frac1{v_{\text{th},\alpha}^2}\partial_t^2\phi_\alpha(q,t)
  &+&q^2\nu_\alpha\partial_t\phi_\alpha(q,t)
  +q^2\phi_\alpha(q,t)\\ \nonumber
  &+&\int_0^t m_\alpha(q,t-t')\partial_{t'}\phi_\alpha(q,t')\,dt'=0\,,
\end{eqnarray}
with initial conditions $\phi_\alpha(q,0)=1$ and $\dot\phi_\alpha(q,0)=0$.
\end{subequations}
We are not interested in the short-time dynamics. Therefore, numerical
solutions are later shown for the overdamped case, where the second derivative
in Eq.~\eqref{phist} and thus inertia terms are dropped.
We can then set $\nu_\alpha=1$ to define
the unit of time. For the resulting Brownian dynamics, the low-density
asymptote of the LG has been worked out exactly \cite{Bauer}.

The memory kernel $m_\alpha(q,t)$
is formally given as a correlation function of generalized fluctuating
forces. MCT assumes the dominant contribution
in the relaxation of these forces to be given by their overlap with
density-pair fluctuations -- for the tagged-particle dynamics, with
pair modes of the form $\varrho_\beta^s(\vec k,t)\varrho_\gamma(\vec p,t)$,
formed with the collective number-density fluctuations
$
  \varrho_\alpha(\vec q,t)=\sum_{k=1}^{N_\alpha}\exp[
    i\vec q\cdot\vec r_{\alpha,k}(t)]
$,
where $\vec r_{\alpha,k}(t)$ is the trajectory of the $k$-th particle
of species $\alpha$. The corresponding collective partial number density
correlation functions are denoted by
$
  \Phi_{\alpha\beta}(q,t)=\left\langle\varrho_\alpha^*(\vec q,t)
  \varrho_\beta(\vec q)\right\rangle
$,
and their initial value is the static structure factor,
$\bs S(q)=\bs\Phi(q,0)$.
The MCT expression for
the tagged-particle memory kernel then reads
\begin{equation}\label{msqt}
  m_\alpha(q,t)\approx
  {\varrho}\int\frac{d^dk}{(2\pi)^d}
  {(\vec e_{\vec q}\vec k)^2}c_{\alpha\alpha'}(k)c_{\alpha\alpha''}(k)
  \Phi_{\alpha'\alpha''}(k,t)\phi_\alpha(p,t)\,.
\end{equation}
where $\varrho$ is the total number density of the $d$-dimensional system.
It is hence expressed as a bilinear functional of the tagged-particle
density correlation function $\phi_\alpha(p,t)$, and of the
collective density correlators $\Phi_{\alpha\beta}(k,t)$. Momentum conservation
implies $\vec q=\vec k+\vec p$, and we denote by $\vec e_{\vec q}$
the unit vector in $\vec q$-direction.
The coupling coefficients are given in terms of the
Ornstein-Zernike direct correlation functions $c_{\alpha\beta}(q)$.

To determine the full dynamics of the system, one would first need to
calculate the collective dynamics, $\bs\Phi(q,t)$, for which analogous
MCT expressions can be derived. In this paper, we are only concerned with
the tagged-particle dynamics, in the case where the slow relaxation time of
the heterogeneous matrix, $\tau_{\alpha\beta}$, is much larger than the time
scales characterizing the nontrivial dynamics of the tracer. We are also
interested in slow dynamics beyond the initial short-time transient,
i.e., in times $1/\nu_\alpha\ll t\ll\tau_{\alpha\beta}$.
In this case, $\bs\Phi(q,t)\approx\bs F(q)\succ\bs 0$, a positive-definite
non-vanishing matrix of nonergodicity factors \cite{Franosch02}.
It is standard MCT procedure to calculate $\bs F(q)$ from the
given static structure factor of the system.
Generically, $\bs F(q)$ decays at large $q$ to provide an implicit
large-$q$ cutoff of the integrals; it is also non-singular at all $q$.
\begin{equation}
  v(k)=\sum_{\alpha\beta\gamma}c_{\alpha\beta}(k)c_{\alpha\gamma}(k)
  F_{\beta\gamma}(k)>0
\end{equation}
is the effective coupling coefficient that is positive and non-singular.
The expression of the MCT memory kernel then reads
\begin{equation}\label{mthroughg}
  m_\alpha(q,t)
  =\varrho\int\frac{d^dk}{(2\pi)^d}{(\vec e_{\vec q}\vec k)^2}
  v(k)\phi_\alpha(p,t)\,.
\end{equation}
Rewriting the integral in terms of bispherical coordinates, i.e., in terms
of moduli $k$ and $p$, one realizes that the $k$-integral can be carried
out immediately, and one is left with a linear functional
\begin{equation}\label{mthroughp}
  m_\alpha(q,t)\equiv m[\phi](q,t)
  =\varrho\int dp\,\tilde v_{qp}\phi_\alpha(p,t)\,,
\end{equation}
where we omit the subscript labeling the species for brevity.

The theory given by Eqs.~\eqref{phist} and \eqref{mthroughg} is a special
case of MCT for the dynamics of a tagged particle in a mixture. In the
presence of quenched rather than self-generated disorder, the derivation
of such a theory becomes more subtle \cite{Krakoviack07}. Nevertheless,
the resulting tagged-particle equations are structurally equivalent
\cite{Krakoviack09}, as is the MCT for the LG, and also the limit of MCT
considering the remaining dynamics of particles deep inside the glass,
the strong-coupling limit \cite{GoetzeBP}.
To simplify the further numerical procedure, we consider
the LG model from now on, i.e., a point tracer among randomly
placed (overlapping) hard-sphere obstacles with radius $R$.
The vertex $v(k)$ can then be evaluated
explicitly; in $d=3$ one obtains
\begin{equation}\label{vlg}
  v_\text{LG}(k)={4(2\pi)^2}R^4(j_1(kR))^2>0\,,
\end{equation}
with $j_1$ the first modified spherical Bessel function.
Also, $\nu=\sqrt\pi\varrho R^2v_{\text{th}}$ in this case.

The nonergodicity parameter $f(q)=\lim_{t\to\infty}\phi(q,t)$ can be
determined \cite{Goetze95b,Franosch02} as the largest positive solution of
\begin{equation}\label{nonergodic}
  \frac{f(q)}{1-f(q)}=m[f](q)/q^2\,.
\end{equation}
It is well known that bifurcations exist in these equations: smooth
changes of the coupling coefficients (such as induced by a smooth increase
in density without structural changes) can lead to singular changes
in $f(q)$.
Standard MCT
considers $m$ to be a
nonlinear functional of the $f$, giving rise to so-called $\mathcal A_\ell$
bifurcations (as classified by Arnol'd) that one identifies
as ideal glass transitions where $m(q)$ jumps to a finite value $m^c(q)>0$
\cite{Goetze.2009}.
In the present context, we are concerned with continuous
transitions, viz.\ the critical points in parameter space, where the
trivial solution, $f(q)=0$ for all finite $q$, changes continuously
to a non-trivial solutions $f(q)\ge0$. At the critical point, $f^c(q)=0$ for
all finite $q$, and consequently $m^c(q)=0$ there. Note that for $q=0$,
$f(0)=1$ holds always. These continuous transitions arise in the LG
upon changing the scatterer density $\varrho$, and in binary-mixture
systems through tuning the parameters such that the combination of
direct correlation functions and nonergodicity parameters entering $v(k)$
leads to sufficiently weak coupling.

Not all continuous MCT transitions are of the
localization-transition type discussed here. For example, in dumbbell-shaped
molecules with top--down symmetry, a glass may form where the
rotational degrees of freedom associated with this symmetry remain ergodic,
and eventually freeze by further compression
\cite{Goetze00,Letz00,Chong02a,Chong02b,Kaemmerer97,Chong05}.
This MCT transition is continuous but does not involve a divergent
static length scale.

If $m[f]$ is a linear functional as in
Eq.~\eqref{mthroughp}, we can express $f$ via Eq.~\eqref{nonergodic} to get
\begin{eqnarray}\label{mfull}
  m(q)&=&\varrho\int dp\,\tilde v_{qp}\frac{m(p)}{p^2+m(p)}\\ \nonumber
  &=&\varrho\int dp\,W_{qp}m(p)-\varrho\int dp\,W_{qp}\frac{m(p)^2}{p^2+m(p)}
  \,.
\end{eqnarray}
a closed nonlinear integral equation for $m(q)$, where we have introduced
$W_{qp}=\tilde v_{qp}/p^2$. Expressing $m$
in this form, one recognizes immediately that, whenever $m^c$ is
zero for finite $k$, the peculiarities of the integral at small $k$ become
important since there is a dangerous denominator.
In the following, we will present both analytical considerations valid
asymptotically close to the transition, and numerical results. The
latter have been obtained by a specially devised numerical grid for
the discretization of integrals like in Eq.~\eqref{mfull}, described
in \ref{numerics}.

\section{Results}\label{asymptotics}

\subsection{Statics}

It will be of further importance to understand the small-wave-number limits
of the vertex $W_{qp}$. This depends on the dimensionality; for $d=3$
one obtains from Eq.~\eqref{mthroughg}
\begin{equation}
  W_{qp}=\frac1{(2\pi)^2}\frac{1}{4q^3p}
  \int_{|q-p|}^{q+p}k\,dk\,(q^2-p^2+k^2)^2 v(k)\,.
\end{equation}
Considering $p\to0$ at fixed $q$,
$
  (2\pi)^2 W_{qp}=2q^2v(q)+\mathcal O(p^2)
$,
and for $q\to0$ at fixed $p$,
$
  (2\pi)^2 W_{qp}={\textstyle\frac23}p\,v(p)+\mathcal O(q^2)
$.
Thus, $W_{qp}$ is regular in the $q\to0$ and $p\to0$ limits, and falls off
quickly enough at large $p$ to ensure the finiteness of the integrals we
consider.

The influence of the small-wavenumber regime in Eq.~\eqref{mfull} is
easiest understood by introducing an infrared cutoff parameter $\delta$
in the integral
and discussing the asymptotic dependence on it. Close
to the transition, $m(q)$ is small; Eq.~\eqref{mfull} becomes a
linear integral equation with positive definite kernel $W_{qp}$.
Nontrivial solutions only exist
if $\varrho$ is an eigenvalue \cite{PorterStirling}.
The equivalent of the Perron-Frobenius theorem for nonnegative matrices holds,
called Jentzsch's theorem in the context of Fredholm integral equations
\cite{Jentzsch}:
there exists a non-degenerate minimal eigenvalue $\varrho_c$ that is positive
and real, and the corresponding eigenfunction can be chosen real and positive.
Obviously, $\varrho_c$ defines the critical density: for densities below,
only the trivial solution $m_\delta(q)=0$ exists, so that the particle
is delocalized. We denote the critical eigenfunction by $h(q)$,
\begin{subequations}
\begin{equation}\label{heq}
  h_\delta(q)=\varrho_c\int_\delta^\infty dp\,W_{qp}h_\delta(p)\,.
\end{equation}
We will also need the eigenfunction $\hat h(q)$ of the transposed equation,
\begin{equation}
  \hat h_\delta(p)=\varrho_c\int_\delta^\infty dq\,\hat h_\delta(q)
  W_{qp}\,.
\end{equation}
\end{subequations}
In order to normalize the eigenfunctions we can impose two conditions,
which we choose for later convenience,
\begin{subequations}
\begin{eqnarray}\label{hnorm}
  \int_\delta^\infty dq\,\hat h_\delta(q)h_\delta(q)=1\,,
&\qquad&
  \hat h_\delta(\delta)h_\delta(\delta)^{3/2}(\pi/2)=1\,.
\end{eqnarray}
\end{subequations}

\begin{figure}
\centerline{\includegraphics[width=\figurewidth]{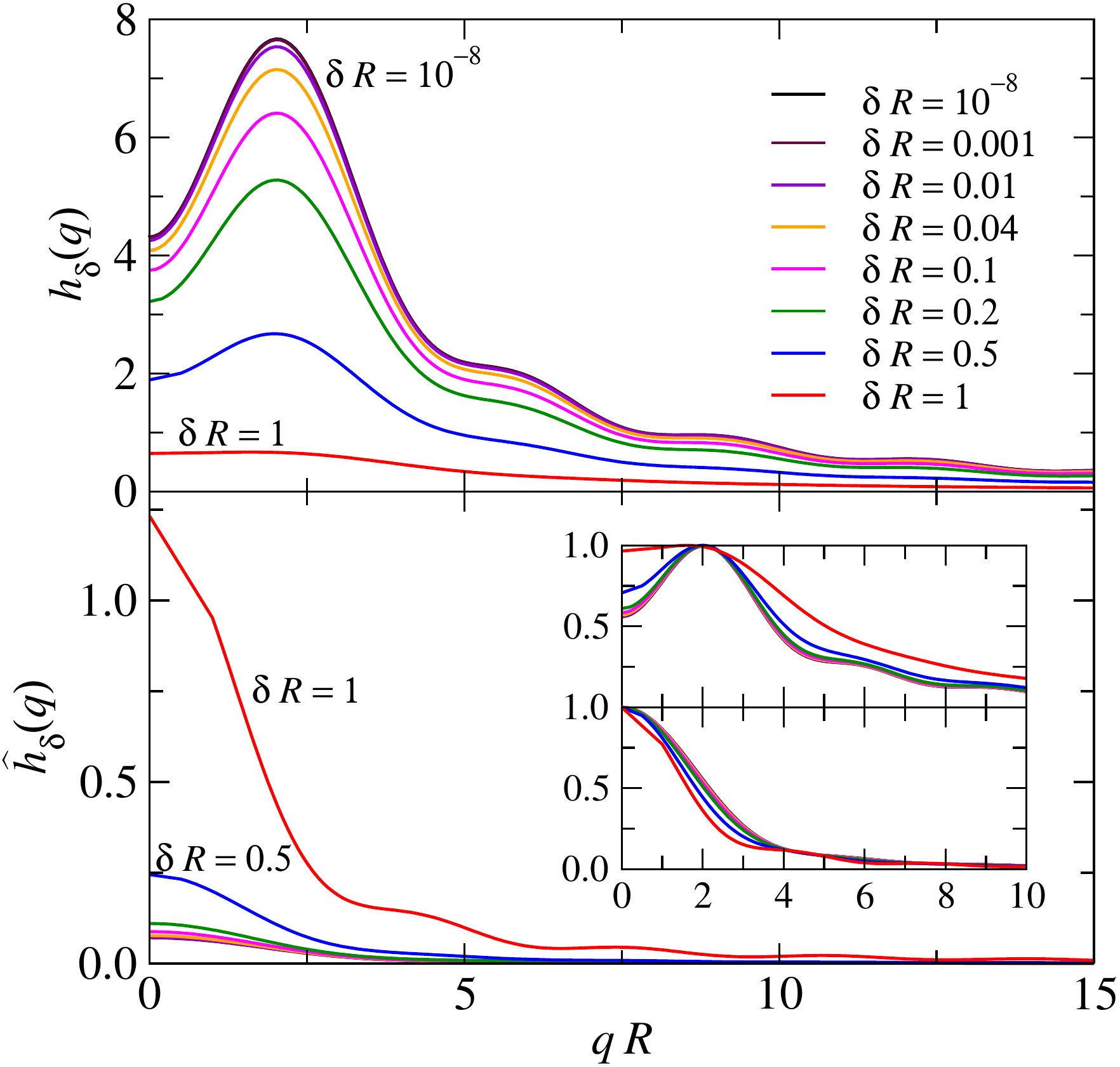}}
\caption{\label{fig:h}
  Critical eigenfunctions $h_\delta(q)$ (upper panel) and $\hat h_\delta(q)$
  (lower panel) for the Lorentz-gas MCT vertex
  for various $\delta$ as indicated, normalized according
  to Eqs.~\eqref{hnorm}. The functions have been
  analytically extended to $q<\delta$. The insets show $h_\delta(q)$
  and $\hat h_\delta(q)$ normalized to their respective maxima.
}
\end{figure}

Critical eigenfunctions $h_\delta(q)$ and $\hat h_\delta(q)$
are shown in Fig.~\ref{fig:h}
for various cutoff values $\delta$, evaluated for the Lorentz-Gas vertex.
Apart from the normalization, the shape of these functions
depends rather insensitively on $\delta$, as exhibited by the insets
of Fig.~\ref{fig:h}.

The next-to-leading order expansion of Eq.~\eqref{mfull} depends rather
more sensitively on $\delta$. We shall distinguish ``large'' cutoff,
$\delta\gg\sqrt{m(\delta)}$, from ``small'' ones,
$\delta\ll\sqrt{m(\delta)}$. In the former case, the term $m(p)$ in
the denominator of the second part of
Eq.~\eqref{mfull} is not relevant close to the transition, and the
corresponding integral yields a term of $\mathcal O(m^2)$. We thus expand
for $\varrho>\varrho_c$
\begin{equation}\label{mexplarge}
  m(q)=\sigma h_\delta(q)+\sigma^2 K_\delta(q)+\mathcal O(\sigma^3)
  \,,
  \qquad\text{$\delta$ large,}
\end{equation}
where $\sigma$ is a small parameter quantifying the distance to the
transition point in a way to be specified. Introducing
$\varepsilon=(\varrho-\varrho_c)/\varrho_c$ and inserting into
Eq.~\eqref{mfull} we get
after multiplication with $\hat h_\delta(q)$ and integration over $q$
the solubility condition
\begin{equation}\label{sigmalarge}
  \sigma=\varepsilon\left(\int_\delta^\infty dq\,
  \hat h_\delta(q)(h_\delta(q)/q)^2\right)^{-1}
  \approx\varepsilon\,{\delta}/({\hat h_\delta(\delta)h_\delta^2(\delta)})
  \,,
  \quad\text{$\delta$ large,}
\end{equation}
where in the approximation we have exploited that the integral is dominated
by contributions from $q\approx\delta$.
In other words, $\sigma\sim\varepsilon$, which is the asymptotic
relation also known from the schematic model \cite{Sjoegren86}.

For small cutoff, $\delta\ll\sqrt{m_\delta(\delta)}$, the second part
in Eq.~\eqref{mfull} needs to be treated with care. Specifically, in the
small-$p$ region we can replace $W_{qp}\approx W_{q\delta}$ and
$m_\delta(p)\approx m_\delta(\delta)$ in order to perform this part
of the integral explicitly. Omitting the remaining terms of
$\mathcal O(m^2)$, we arrive at
\begin{equation}\label{mexpand}
  m(p)=\varrho\int_\delta^\infty dp\,W_{qp}m(p)
  -\varrho W_{q\delta}m^{3/2}(\delta)\frac{\pi}{2}
  +{\mathcal O}(m^2)\,.
\end{equation}
We now need to expand differently. Again for $\varrho>\varrho_c$,
\begin{equation}\label{mexpsmall}
  m(q)=\sigma h_\delta(q)+\sigma^{3/2}\tilde K_\delta(q)
  +\mathcal O(\sigma^2)\,,
  \qquad\text{$\delta$ small.}
\end{equation}
Inserting this into Eq.~\eqref{mexpand} and collecting terms of
$\mathcal O(\sigma^{3/2})$, we obtain
\begin{eqnarray}
  \tilde K_\delta(q)&=&\varrho_c\int dp\,W_{qp}\tilde K_\delta(p)\\
  &+&\sigma^{-1/2}\varepsilon\varrho_c\int dp\,W_{qp}h_\delta(p)
  -\varrho_c W_{q\delta}h_\delta^{3/2}(\delta)(\pi/2)\,,
\end{eqnarray}
if we set $\varepsilon\sim\sigma^{1/2}$.
The solubility condition is again obtained by contracting with
$\hat h_\delta(q)$ and reads
\begin{equation}\label{sigmasmall}
  \sqrt{|\sigma|}=|\varepsilon|\big/
  \hat h_\delta(\delta)h_\delta(\delta)^{3/2}(\pi/2)\,,
  \qquad\text{$\delta$ small,}
\end{equation}
where it is understood that $\varepsilon\gtrless0$ corresponds to
$\sigma\gtrless0$ and denotes the localized respective delocalized side
of the transition.

Thus, the asymptotic behavior of $m_\delta(q)$ changes
depending on $\delta$. Stated differently, for any fixed infrared cutoff
there exist regimes close to the transition where the ``large-$\delta$''
asymptote is reached. For distances $\varepsilon$ larger than a cross-over
distance, the ``small-$\delta$'' asymptote holds. We estimate this
cross-over by equating the leading-order terms, from which one gets
\begin{equation}\label{epsilonstar}
  \varepsilon^*=\frac{\hat h_\delta(\delta)^2h_\delta(\delta)^3(\pi/2)^2}
  {\int_\delta^\infty dq\,\hat h_\delta(q)h_\delta(q)^2/q^2}
  \approx \hat h_\delta(\delta)h_\delta(\delta)(\pi/2)^2\delta\,.
\end{equation}

\begin{figure}
\centerline{\includegraphics[width=\figurewidth]{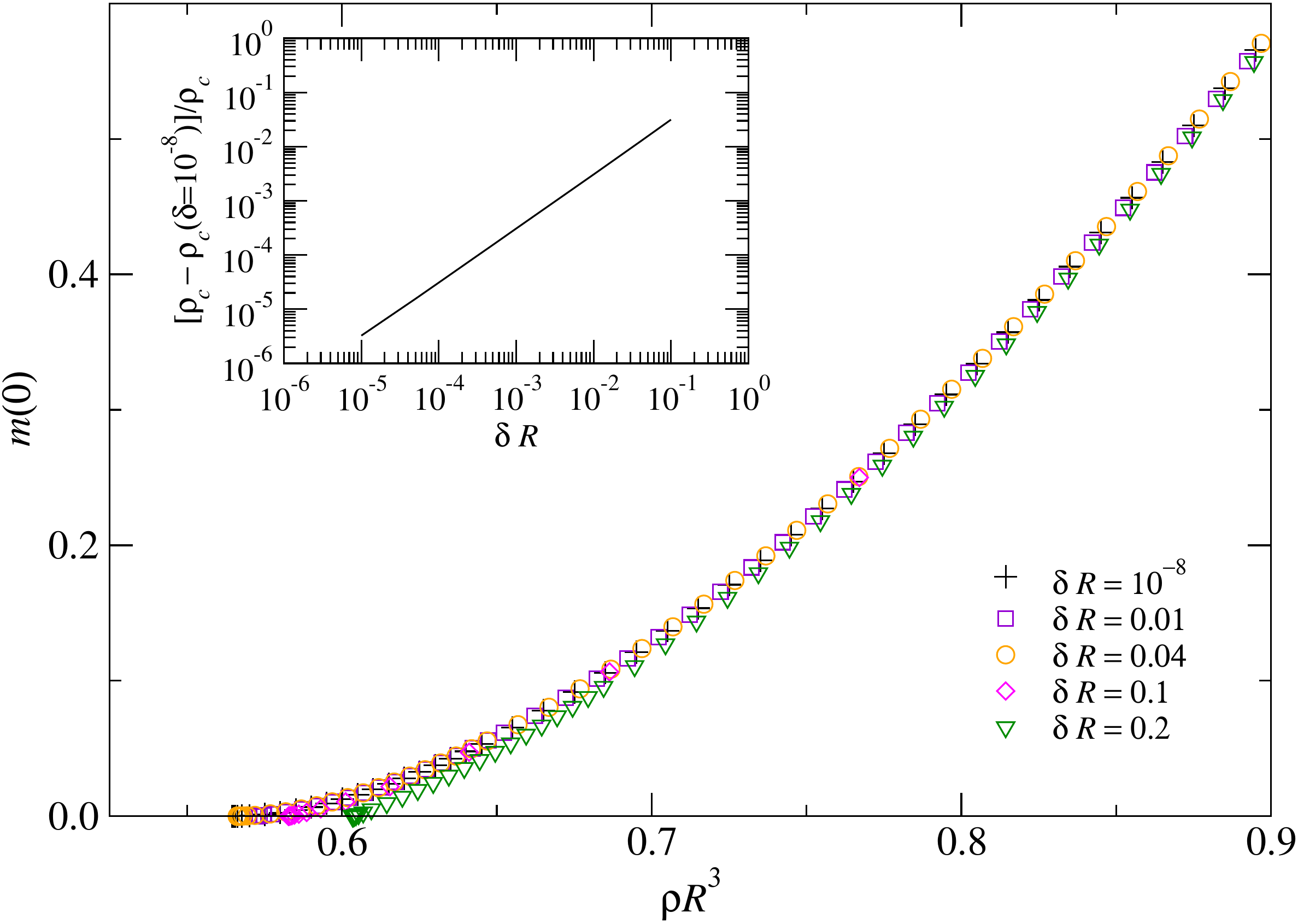}}
\caption{\label{fig:m}
  Order parameter $m(0)$ as a function of density $\varrho$,
  for various $\delta$ as indicated. The inset shows the variation
  of the critical point, $\varrho_c$, with $\delta$.
}
\end{figure}

To demonstrate this behavior, let us investigate the order parameter
$m(0)$ as a function of density. Numerical results are shown in
Fig.~\ref{fig:m}.
Except in a region of densities around $0.6$, the curves
depend relatively weakly on $\delta$, indicating that the theory
in $d=3$ is for this aspect not
sensitive to an infrared cutoff sufficiently far in the localized region.
A numerical determination of the critical eigenvalue
in Eq.~\eqref{heq} for the LG vertex gives $\varrho_c\approx0.5644/R^3$
for $\delta\to0$.
As the inset of Fig.~\ref{fig:m} demonstrates, for the
$d=3$ LG vertex, the variation of $\varrho_c$ with $\delta$ is
small.
The localization point found in simulations
of the three-dimensional Lorentz gas is $\varrho_c\approx0.84/R^3$
\cite{Hoefling06,Hoefling08}.
It is well known that MCT over-estimates the tendency to arrest,
so that we will not be concerned by this discrepancy.
Note that a generalized hydrodynamic approach, replacing all occurrences
of the memory kernel in Eq.~\eqref{mfull} by its $q\to0$ limit,
allows to directly evaluate the integral using Eq.~\eqref{vlg} \cite{Goetze81b},
resulting in $\varrho_c=9/(4\pi R^3)\approx0.716R^{-3}$.

\begin{figure}
\centerline{\includegraphics[width=\figurewidth]{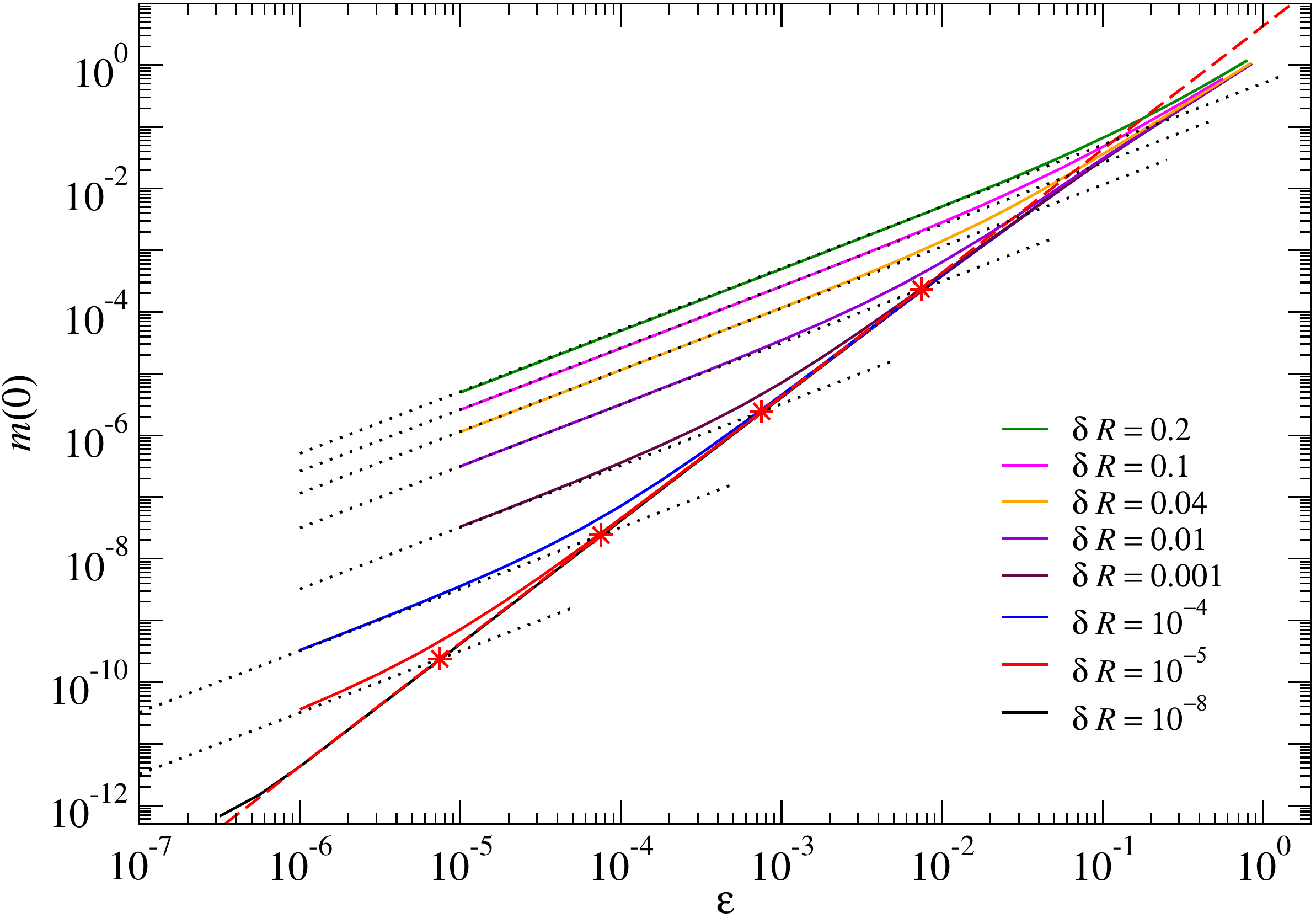}}
\caption{\label{fig:mlog}
  Same as Fig.~\ref{fig:m}, but plotted double-logarithmically as a
  function of the distance to the transition
  $\varepsilon=(\varrho-\varrho_c)/\varrho_c$.
}
\end{figure}

The different asymptotic regimes are exhibited in Fig.~\ref{fig:mlog},
where $m(0)$ is shown as a function of $\varepsilon$ in
double-logarithmic fashion. The asymptotes, Eq.~\eqref{mexplarge} together
with Eq.~\eqref{sigmalarge}, and Eq.~\eqref{mexpsmall} with
Eq.~\eqref{sigmasmall}, are shown as dotted lines. They are found to
agree well with the numerical results. Stars in Fig.~\ref{fig:mlog}
mark the cross-over point $\varepsilon^*$, Eq.~\eqref{epsilonstar},
where the two asymptotic regimes meet.

\begin{figure}
\centerline{\includegraphics[width=\figurewidth]{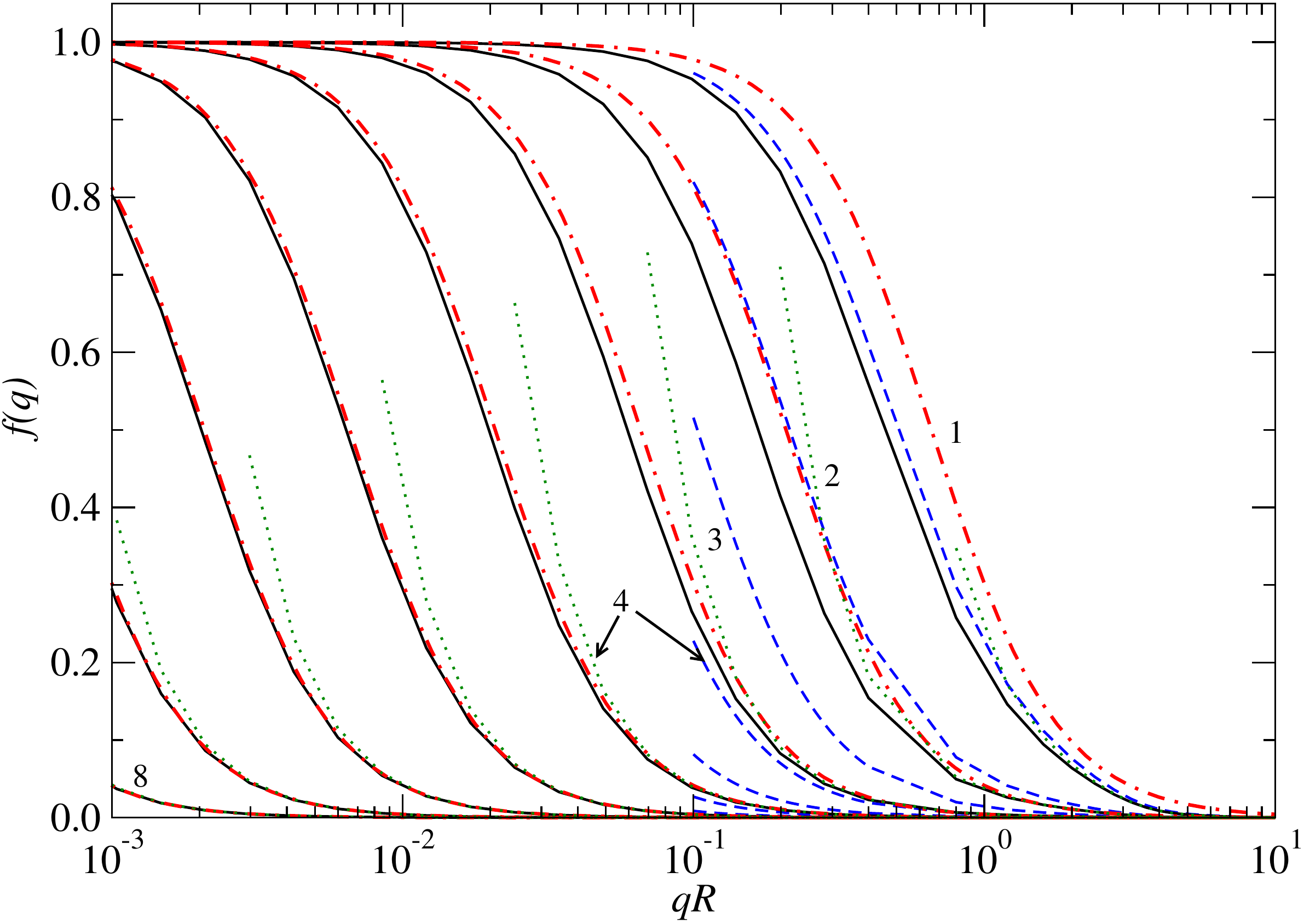}}
\caption{\label{fig:fq}
  Nonergodicity parameter $f(q)$ corresponding to $\delta=10^{-8}$
  (solid lines) and $\delta=0.01$ (dashed lines), for $\varepsilon=10^{-n/2}$
  as labeled. The asymptotic results
  given $m(q)/q^2$ and Eq.~\eqref{fqasy2} are shown in
  dotted and dash-dotted.
}
\end{figure}

Figure~\ref{fig:fq} shows the nonergodicity parameters $f(q)$ obtained
for $\delta=10^{-8}$ and $\delta=0.01$ for various $\varepsilon$.
Inserting the leading-order result $m(q)=|\sigma|h_\delta(q)$ into
Eq.~\eqref{nonergodic}, we get for $\sigma>0$
\begin{equation}\label{fqasy2}
  f(q)\sim\frac{1}{1+q^2/(\sigma h_\delta(q))}\,,
\end{equation}
so that for small $q$, $f(q)$ becomes a Lorentzian
from which we read off a length scale diverging as $\sim\sigma^{-1/2}$.

\subsection{Dynamical Scaling Functions}

We are now interested in the dynamic scaling of correlation functions
in the limit of large times, i.e., time scales much longer than that of
binary collisions. We shall thus analyze Eq.~\eqref{phisz} in the limit
$|z|\ll\nu$.
Analogously to Eq.~\eqref{mfull}, we get
\begin{equation}\label{mszfull}
  zm(q,z)=\varrho\int_\delta^\infty dp\,W_{qp}z\mu(p,z)\\
  +\varrho\int_\delta^\infty dp\,W_{qp}\frac{(z\mu(p,z))^2}
  {p^2-z\mu(p,z)}\,,
\end{equation}
where $z\mu(p,z)=z^2/v_{\text{th}}^2+p^2i\nu z
+zm(p,z)$. Note in passing that the emerging integrals are well-defined
if one assumes $\phi(q,t)$ and $m(q,t)$ to be
completely monotone functions. In particular, there then holds
for the real part
$\Re m(q,z)\ge0$ for $\Re z<0$ \cite{Gripenberg}, ensuring the
analog of Eq.~\eqref{mexpand} with $m$ replaced by $\mu$ on the r.h.s.
One readily obtains that in leading order at the critical point, $m(q,z)$
solves the eigenvalue equation \eqref{heq} as $z\to0$. Thus,
$m(q,z)=h_\delta(q)G(z)+\delta m(q,z)$. The scaling
ansatz observing $m(q)\sim\mathcal O(|\sigma|)$ sets
$G(z)=|\sigma|t_\sigma g(\hat z)$
for rescaled frequencies $\hat z=zt_\sigma$. Here $t_\sigma$ is a
time scale that diverges faster than $1/|\sigma|$ as the transition
is approached. The leading-order term is reminiscent of
the standard MCT factorization theorem \cite{Goetze.2009}: wave-number- and
frequency-dependence factorize. For the correlator this implies
$\phi(q,t)\sim(h(q)/q^2)G(t)$ as $\sigma\to0$. But since $\phi(q,t)\le1$,
at the localization transition the factorization theorem for $\phi(q,t)$
is violated even as $t\to\infty$ for small $q$. This is demonstrated
in Fig.~\ref{fig:fq}, where dotted lines show
$f(q)\sim\sigma h(q)/q^2\sim m(q)/q^2$.

\begin{figure}
\centerline{\includegraphics[width=\figurewidth]{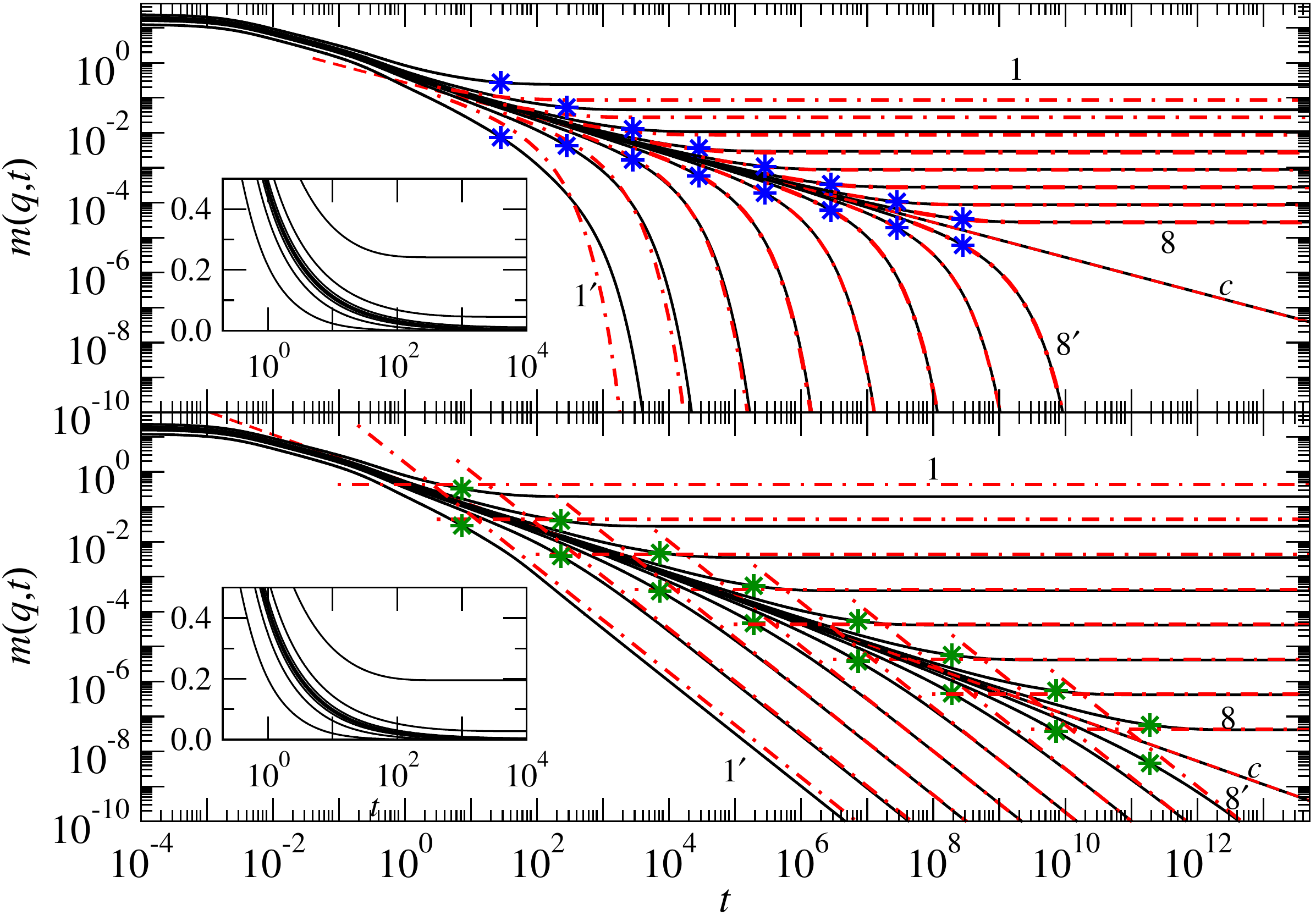}}
\caption{\label{fig:mz}
  Memory kernel $m(q,t)$ at $q=\delta$, for the case of large $\delta$
  (top panel) and small $\delta$ (bottom panel), for distances to the
  respective transition points $\varepsilon=\pm10^{-n/2}$ ($n=1,\ldots 8$)
  and $\varepsilon=0$ (curves labeled $c$). The
  asymptotic results obtained from the scaling equations,
  Eqs.~\eqref{scalinglarge} and \eqref{scalingsmall}, are shown as
  dash-dotted lines.
  Stars mark the time scales $t_\sigma$ used in the scaling equation.
  Insets: $m(q,t)$ on a linear--log scale.
}
\end{figure}

To obtain the scaling equation, we need to consider the
next-to-leading terms and hence distinguish again between small and large
$\delta$.
In the latter regime, we set, in generalization of Eq.~\eqref{mexplarge},
\begin{equation}\label{mzexplarge}
  m(q,z)=|\sigma|t_\sigma h_\delta(q)g(\hat z)\\
  +\sigma^2t_\sigma\hat K(q,\hat z)+\mathcal O(|\sigma|^3)\,,
  \qquad\text{$\delta$ large.}
\end{equation}
Inserting into Eq.~\eqref{mszfull} and balancing the terms of
$\mathcal O(\sigma^2)$ yields, after contraction
with the left-eigenfunction $\hat h_\delta(q)$, the scaling equation
$
  i(\tau/t_\sigma)\hat z+\varepsilon|\sigma|\hat zg(\hat z)
  +\varepsilon\sigma(\hat zg(\hat z))^2=0
$.
Here we have used Eqs.~\eqref{hnorm} and \eqref{sigmalarge} and defined
$\tau=\nu\int dp\,\hat h_\delta(p)p^2$. We thus fix
\begin{equation}\label{tsigma}
  t_\sigma=\frac{\tau}{|\varepsilon||\sigma|}\,,
\end{equation}
and get the two scaling solutions for $\sigma\gtrless0$,
\begin{equation}\label{scalinglarge}
  0=\pm i+g_\pm(\hat z)\pm\hat z(g_\pm(\hat z))^2\,,
  \qquad\text{$\delta$ large.}
\end{equation}
This is the scaling equation derived earlier by G\"otze for the F1 model
\cite{Goetze.2009}. Its solution can be given as a closed expression for the
inverse Laplace transform, $g(\hat t)$, where $\hat t=t/t_\sigma$.
Here we are only interested in the asymptotes. Letting $\hat z\to0$,
we find $g_+(\hat z)\sim1/\hat z$. Thus, $g_+(\hat t)\to1$ for $\varepsilon>0$,
i.e., the signature of localized dynamics. On the other hand $g_-(\hat z)$
remains regular, indicating that $g_-(\hat t)$ decays faster than $1/\hat t$
at long times (exponential in fact). Letting $\hat z\to\infty$ we obtain
$g(\hat t)\sim(\pi\hat t)^{-1/2}$, so that for the time window
$1/\nu\ll t\ll t_\sigma$, we have the $\sigma$-independent asymptote
\begin{equation}\label{mpowerlarge}
  m(q,t)\sim|\sigma|\frac{h_\delta(q)}{\sqrt{\pi}}\left(\frac{t}{t_\sigma}
  \right)^{-1/2}
  \sim
  \frac\delta{\sqrt{\pi h_\delta(\delta)}}\left(\frac{t}{\tau}\right)^{-1/2}\,,
  \quad\text{$\delta$ large.}
\end{equation}
Note that an explicit dependence on the cutoff parameter remains in the
prefactor.
The top panel of Fig.~\ref{fig:mz} exemplifies $m(q\!=\!\delta,t)$
in the large-$\delta$ regime for various distances $\varepsilon$ to
the transition. Dash-dotted
lines indicate the above asymptotes.

Along similar lines, we deduce the small-$\delta$ scaling function.
Here we have to expand as in Eq.~\eqref{mexpsmall},
\begin{equation}
  m(q,z)=|\sigma|t_\sigma h_\delta(q)g(\hat z)
  +|\sigma|^{3/2}t_\sigma\tilde K(q,\hat z)\,,
  \qquad\text{$\delta$ small.}
\end{equation}
Balancing the $|\sigma|^{3/2}$ terms in Eq.~\eqref{mszfull}
we obtain after multiplying with $\hat h_\delta(q)$ and integration over $q$
\begin{equation}
  i(\tau/t_\sigma)\hat z+\varepsilon|\sigma|\hat zg(\hat z)
  +\hat h_\delta(\delta)h_\delta(\delta)^{3/2}(\pi/2)|\sigma|^{3/2}
  (\hat zg(\hat z))^{3/2}=0\,.
\end{equation}
Now, using Eqs.~\eqref{sigmasmall} and \eqref{tsigma}, this yields
for $\varepsilon\gtrless0$
\begin{equation}\label{scalingsmall}
  0=\hat z(g_\pm(\hat z))^3+(g_\pm(\hat z))^2
  \pm2ig_\pm(\hat z)-1\,,
  \qquad\text{$\delta$ small.}
\end{equation}
As $\hat z\to\infty$, this equation is solved by
$g_\pm(\hat z)=i(i\hat z)^{-1/3}$,
and thus $g_\pm(\hat t)\sim\hat t^{-2/3}/\Gamma(1/3)$ for
$1/\nu\ll t\ll t_\sigma$.
As $\hat z\to0$, we note again $g_+(\hat z)\sim-1/\hat z$ (localized solution),
but $g_-(\hat z)\sim i-\sqrt{i\hat z}$ leading to
$g_-(\hat t)\sim\hat t^{-3/2}/\sqrt{4\pi}$.
The behavior we find is thus
\begin{eqnarray}\label{mpowersmall}
  m(q,t)\sim (\hat h_\delta(\delta)\pi/2)^{-2/3}/\Gamma(1/3)\,
  (t/\tau)^{-2/3}\,,
  \quad\text{$t\ll t_\sigma$, $\delta$ small,}\\
  m(q,t)\sim|\sigma|h_\delta(q)/\sqrt{4\pi}\,(t/t_\sigma)^{-3/2}\,,
  \quad\text{$t\gg t_\sigma$, $\varepsilon\to0^-$, $\delta$ small.}
\end{eqnarray}
These power laws are exemplified in the lower panel of Fig.~\ref{fig:mz},
for various $\varepsilon$ in the small-$\delta$ regime.
As for the large-$\delta$ regime (upper panel), the asymptotic results
describe the numerical solution for $m(q,t)$ well even for
$\varepsilon=\mathcal O(0.1)$.
Treating the small-wave-number regime of the vertex seriously, a
long-time tail for $m(q,t)$ results; however, this is not the
known universal long-time tail of the LG \cite{Beijeren,Bauer}
in $d=3$, $t^{-5/2}$, which is not contained in the present MCT.

\begin{figure}
\centerline{\includegraphics[width=\figurewidth]{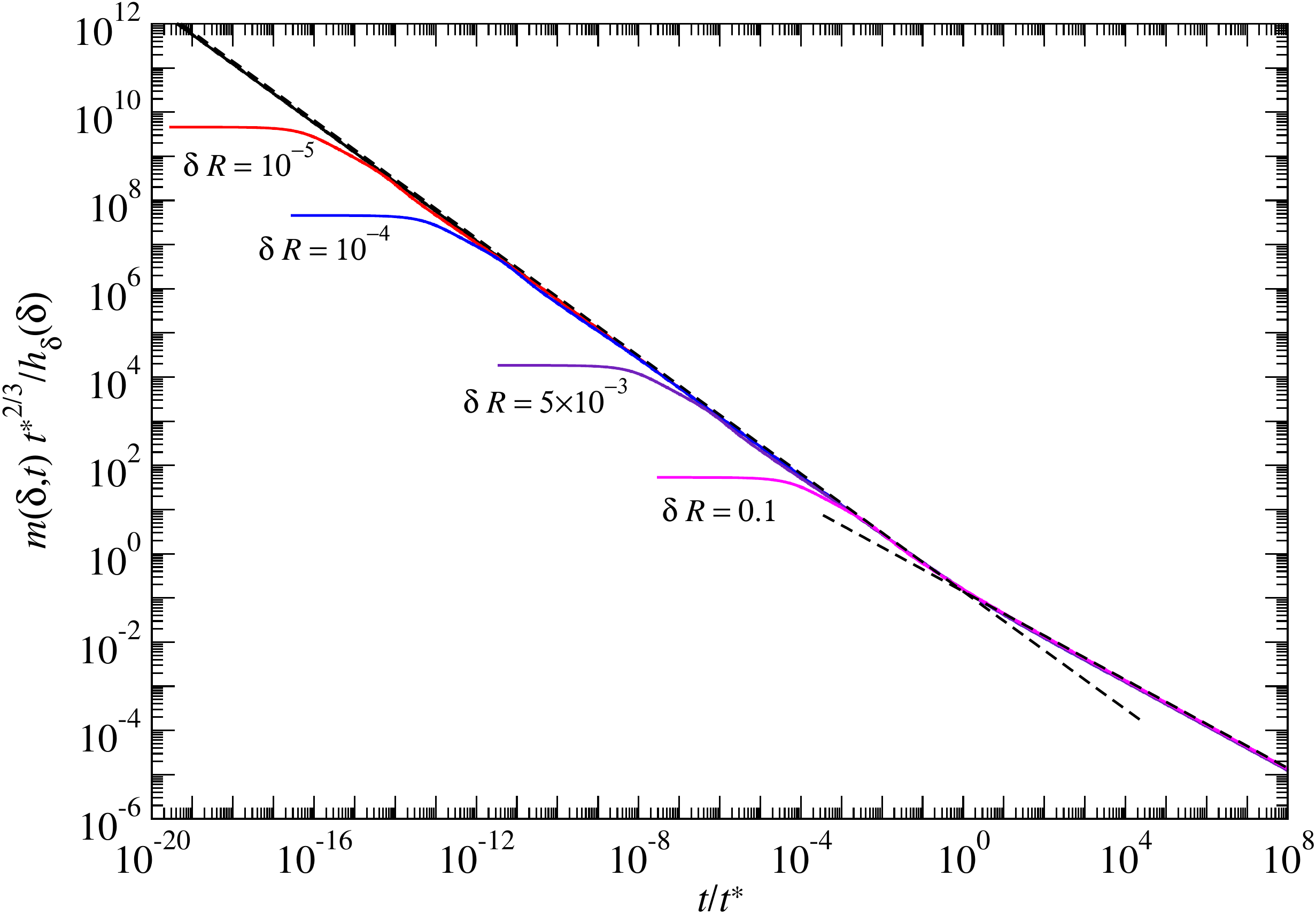}}
\caption{\label{fig:mcoscaled}
  $m(\delta,t)$ at the critical point as a function of $t/t^*$
  according to Eq.~\eqref{tstar}, for various $\delta$ as indicated.
  For the unlabeled curve, $\delta=10^{-8}/R$.
}
\end{figure}

The two regimes outlined above, large and small $\delta$ are separated
by the condition $\delta\sim\sqrt{|zm(\delta,z)|}$. For fixed $\delta$,
this implies a time scale $t^*$ where the small-$\delta$ asymptotics for
$t\ll t^*$ crosses over to the large-$\delta$ asymptotics for $t\gg t^*$.
Equating Eqs.~\eqref{mpowerlarge} and \eqref{mpowersmall},
and using the approximation Eq.~\eqref{epsilonstar}, we find
\begin{equation}\label{tstar}
  t^*\approx\frac{16}{\pi\Gamma(1/3)^6\hat h_\delta(\delta)}\frac{\tau}{\delta^3}\,.
\end{equation}
Thus, plotting $m(\delta,t)$ at various $\delta$ as a function of
$\bar t=t/t^*$ reveals a master curve with asymptotes $\bar t^{-2/3}$
for $\bar t\ll1$, and $\bar t^{-1/2}$ for $\bar t\gg1$, for $\sigma\to0$;
see Fig.~\ref{fig:mcoscaled}.

\begin{figure}
\centerline{\includegraphics[width=\figurewidth]{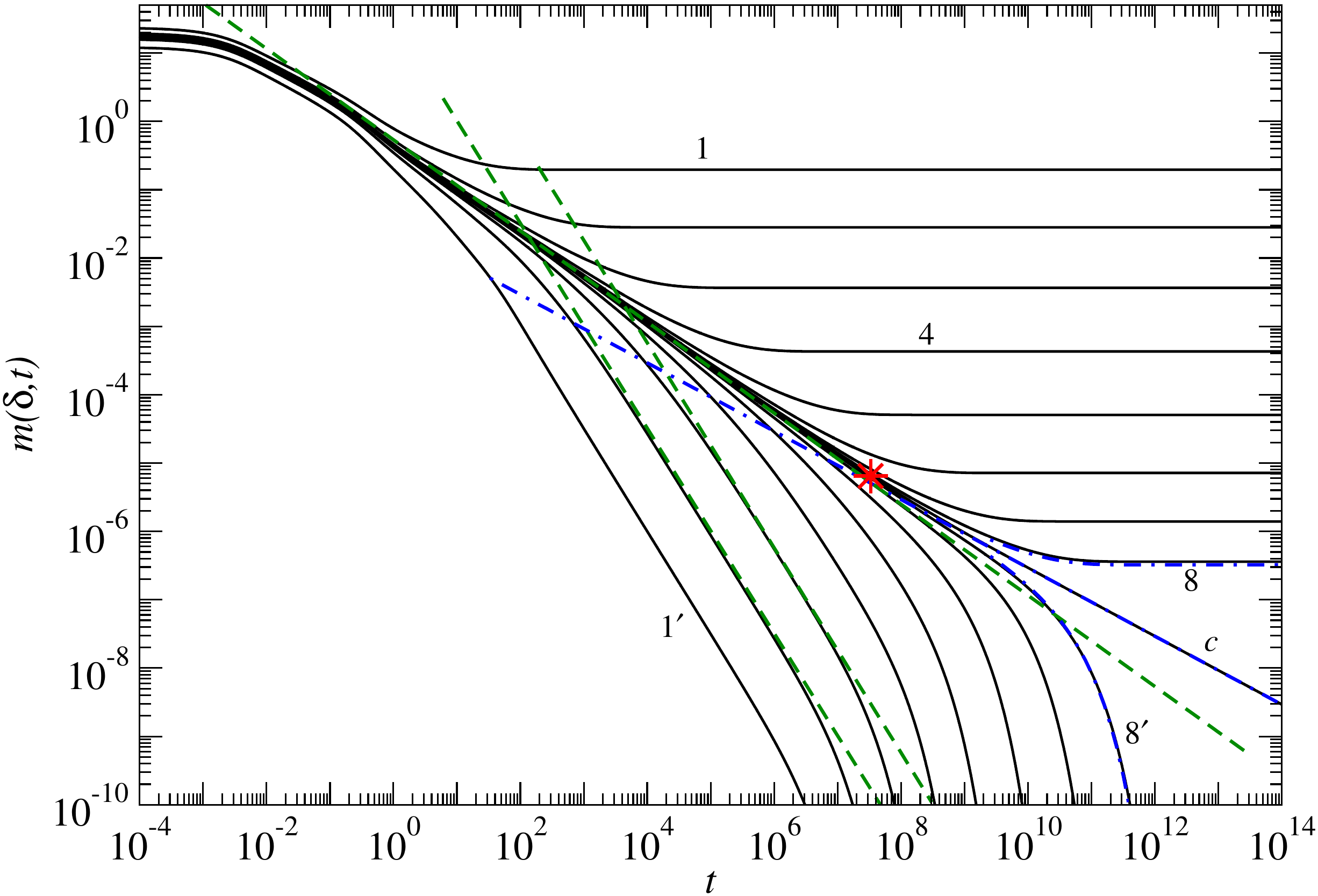}}
\caption{\label{fig:md}
  $m(\delta,t)$ as a function of $t$, for $\delta=10^{-3}$, at various
  distances to the transition, $\varepsilon=\pm10^{-n/2}$ as labeled
  and $\varepsilon=0$ (label $c$).
  Dashed lines show small-$\delta$ asymptotes, dash-dotted lines the
  large-$\delta$ asymptotes at the critical point and for exemplary
  $\varepsilon$. A star marks $t^*$, Eq.~\eqref{tstar}.
}
\end{figure}

If the cutoff is held fixed, the time scale $t_\sigma$ marks the
cross-over between the different power-law regimes.
The small-$\delta$ asymptotes are always encountered first, and if
the corresponding $t_\sigma<t^*$, the cutoff-influenced large-$\delta$
asymptote is not seen. Only close enough to the transition does the
cutoff-dependent small-$\delta$ asymptote become evident for long times.
This can be seen in Fig.~\ref{fig:md}, where a star marks
$t^*$, and dashed and dash-dotted lines the different asymptotic
solutions.

\subsection{Mean-Squared Displacement}

The mean-squared displacement $\delta r^2(t)$ is obtained from the
$q\to0$ limit of the correlator, $\phi(q,t)=1-(q^2/6)\delta r^2(t)$.
Inserting into Eq.~\eqref{phist}, one gets after Laplace-transforming
the equivalent of Eq.~\eqref{phisz},
\begin{equation}\label{msdz}
  \delta r^2(z)=\frac1{z^2}\frac{6v_\text{th}^2}{z+i\nu+m(0,z)}
  \sim\frac{6v_\text{th}^2}{|\sigma|h_\delta(0)t_\sigma}
  \frac1{{\hat z}^2g(\hat z)}\,.
\end{equation}
The latter asymptotic behavior is seen from inserting the leading-order
result for $m(0,z)$ and noticing that $t_\sigma\to\infty$ close to
the transition.
The generic features of $\delta r^2(t)$ at long times are then immediately
clear from those of $m(0,t)$.
For $\varepsilon>0$, from $g(\hat z)\sim-1/\hat z$ there results the
localization length
\begin{equation}\label{loclen}
  r_c=\sqrt{\delta r^2(\infty)/6}=\frac{v_\text{th}}{|\sigma|^{1/2}h_\delta^{1/2}(0)}\,,
  \qquad\text{$\varepsilon>0$.}
\end{equation}
On the liquid side,
$g_-(\hat z)$ remains regular as $\hat z\to0$. We then get
ordinary diffusion,
\begin{equation}\label{diff}
  \delta r^2(t\to\infty)\sim\frac{6v_\text{th}^2}{|\sigma|h_\delta(0)}
  \frac{t}{t_\sigma}\,,
  \qquad\text{$\varepsilon<0$.}
\end{equation}
In the intermediate time window, $1/\nu\ll t\ll t_\sigma$, the power
laws identified for the memory kernel, $m(q,t)\sim t^x$ with $0<x<1$,
result in power laws $\delta r^2(t)\sim t^x$. We get
\begin{equation}\label{msdpower}
  \delta r^2(t)\sim
  \left(\frac{t}{\tau}\right)^x\,,
\end{equation}
an intermediate window of anomalous power-law diffusion,
with exponents $x=1/2$ in the large-$\delta$ case, and $x=2/3$ in the
case of small $\delta$.

\begin{figure}
\centerline{\includegraphics[width=\figurewidth]{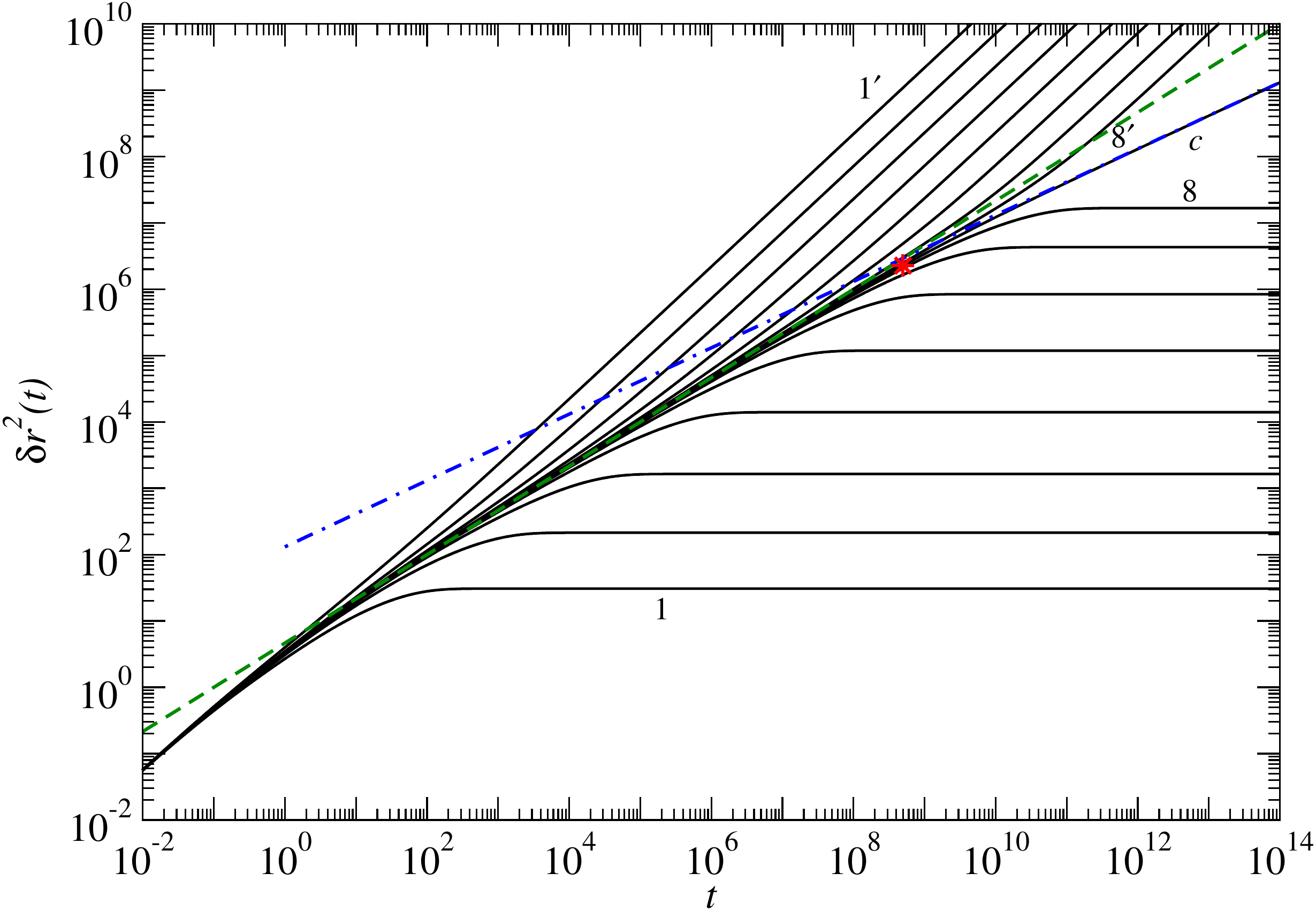}}
\caption{\label{fig:msd}
  Mean-squared displacement $\delta r^2(t)$ vs.\ $t$ for $\delta=10^{-3}$
  and various $\varepsilon$ corresponding to the curves in
  Fig.~\ref{fig:md}. The anomalous-diffusion asymptote, $t^{1/2}$
  ($t^{2/3}$) is shown as a dash-dotted (dotted) line. A star marks
  $\bar t^*$, Eq.~\eqref{tstarmsd}.
}
\end{figure}

Figure~\ref{fig:msd} exemplifies the behavior for $\delta=10^{-3}$,
corresponding to the memory kernels shown in Fig.~\ref{fig:md}. As
stated above, the results are for overdamped Brownian dynamics; this
version of Eq.~\eqref{msdz} reads
$\delta r^2(z)=(6/z^2)/[i\nu+m(0,z)]$, and the relevant asymptotic
results are obtained by replacing $v_\text{th}^2=1$ in
Eqs.~\eqref{loclen} to \eqref{msdpower}.
Again,
a time scale $\bar t^*$ can be identified where the contributions from
the two asymptotic solutions balance. As in Eq.~\eqref{tstar},
$\bar t^*/\tau\sim1/\delta^3$, but the prefactors are different:
\begin{equation}\label{tstarmsd}
  \bar t^*\approx\frac2\pi\left(\frac{2\sqrt2\Gamma(5/3)}{\pi}\right)^6
  \frac{\tau}{\delta^3}\,.
\end{equation}
Note that $\bar t^*\approx 10t^*$, as is verified by comparing the
marked positions in Fig.~\ref{fig:md} and \ref{fig:msd}.

Recall that depending on $\delta$ the connection between $\varepsilon$
and $\sigma$ changes. However, $|\sigma|t_\sigma=\tau/|\varepsilon|$,
and $h_\delta(0)$ changes relatively weakly with $\delta$,
so that Eq.~\eqref{diff} yields the same diffusion law, $\delta r^2(t)=6Dt$
irrespective of $\delta$, with a diffusion coefficient $D\sim|\varepsilon|$.
On the localized side of the transition, Eq.~\eqref{loclen} yields
$r_c\sim\varepsilon^{-1}$ for the small-$\delta$ case, but
$r_c\sim\varepsilon^{-1/2}$ for the large-$\delta$ case.

\section{Conclusions}\label{conclusions}

We have discussed the different asymptotic regimes that emerge from MCT
for the tagged-particle dynamics close to a localization transition,
depending on an arbitrarily chosen infrared cutoff parameter $\delta$.
Such a parameter is not justified a~priori, but can be rationalized
by recognizing that the small-$q$ structure of the MCT vertex for the
LG case studied here is
erroneous \cite{Leutheusser1983}.
Reassuringly, in $d=3$, the predicted
position of the localization transition depends only weakly on $\delta$,
although its overall value suffers from similar disagreement with
computer-simulation result as the MCT-predicted glass transition point.
Note however that in $d=2$ the situation changes. Following an argument
of Ref.~\cite{Goetze79}, one can show that the present MCT does not allow
solutions recovering long-time diffusion if $\delta\to0$; this results
in a transition point that shifts to zero with decreasing $\delta$.
An analysis like the one presented here can still be carried through, if
the changed structure of the vertex $\tilde v_{qp}$ is taken
into account \cite{Schnyder}.

Depending on the distance to the transition $\varepsilon$, and the
cutoff $\delta$, one observes power laws with different exponents for
the memory kernel $m(q,t)$; one distinguishes the small-$\delta$ dynamics
from the large-$\delta$ case, separated by a time scale $t^*$.
This also holds for other dynamic quantities, like the mean-squared
displacement $\delta r^2(t)$. The latter grows as $t^{1/2}$ for
the large-$\delta$ case, which is the result also known from schematic MCT
\cite{Sjoegren86} and is the relevant result for previous numerical
MCT calculations that do not treat the small-$q$ anomalies of the vertex
seriously. It is also, together with the $D\sim|\varepsilon|$ obtained
above, the result expected from mean-field theory
\cite{Stauffer}.
The results presented here also serve the purpose to highlight which parameter
regimes can safely be treated by the present MCT in its widely used
incarnation on a discrete wave-vector grid (where typically,
$\delta\approx0.2/R$), at least in $d=3$.


Our introduction of a cutoff has analogies to a
modern formulation of the renormalization group (RG) approach
\cite{Berges:2002},
where an infra-red cutoff controls the RG flow of the effective
average action.
It would be promising to establish an RG flow for the MCT equations,
possibly
following along the lines of a cluster-MCT \cite{CatesKroy}, although
it is not obvious how to do this.

\ack{
This work was funded in part by the Deutsche Forschungsgemeinschaft,
DFG, project P8 of the Research Unit FOR1394 ``Nonlinear Response to
Probe Vitrification''.
Th.~V.\ thanks for funding through the Helmholtz-Forschungsgemeinschaft
(Impuls- und Vernetzungsfonds, VH-NG-406) and
Zukunftskolleg, Universit\"at Konstanz.
}

\begin{appendix}
\section{Numerical Procedure}\label{numerics}

We sketch the numerical scheme used to solve Eq.~\eqref{mfull}
and similar integral equations. To capture the behavior at small wave
numbers, a logarithmic grid was introduced, containing $N_s$ points
starting at $\delta$ up to some $\Delta q$; from $\Delta q$ to an
upper cutoff $Q$, $N_Q$ equidistantly spaced points complete the grid.
We used $N_s=50$, $N_Q=300$, $\Delta q=0.08/R$, and $Q=48/R$
for static calcuations. Time-dependent quantities were obtained
with $\Delta q=0.4$ and $Q=24/R$. These discretizations were
found to be sufficiently accurate and have been checked with
larger grids in some cases \cite{Schnyder}.

The critical eigenfunctions $h_\delta(q)$ and $\hat h_\delta(q)$ have
been obtained by iteratively solving Eq.~\eqref{mfull} starting with
some constant initial guess; at each iteration step the approximate solution
is inserted in the r.h.s., and the result is taken as the next approximant.
As the iteration convergence is dominated by the eigenvalue $\varrho_c$,
this procedure also determines the critical density, calculated as the
ratio between the results of two iteration steps at fixed $q$.

For the time-dependent Eq.~\eqref{phist}, a standard MCT algorithm is used
where an equidistant grid is chosen for the time $t$, which is repeatedly
coarsened by a factor $2$ after a fixed number of steps in order to cover
the required number of decades.

\end{appendix}

\bibliographystyle{iopart-num}
\bibliography{lit}
\end{document}